\documentclass{article} 
\usepackage{jamaica04}
\usepackage{graphicx}
\frompage{000} \topage{000}                                              
\def\rightmark{multiparticle collisions in a parton cascade}
\def\leftmark{C.Greiner and Z.Xu}
 
\title{Inelastic multiparticle collisions in a parton cascade}

\authors{
{C. Greiner $^{1}$ and Z. Xu $^{1,2}$ %
}\\[2.812mm]
{\normalsize 
$^1$Institut f\"ur Theoretische Physik der Johann Wolfgang Goethe
Universit\"at, \\
\phantom{$^1$}D-60054, Frankfurt am Main, Germany } \\
{\normalsize 
$^2$Institut f\"ur Theoretische Physik der Justus Liebig
Universit\"at, \\
\phantom{$^1$}D-35392, Giessen, Germany
\\[0.2ex] 
}}
 
\abstract{We elaborate on a new 3+1 dimensional 
Monte Carlo parton cascade 
solving kinetic  Boltzmann processes including 
- for the first time - inelastic multiplication
processes ($gg\leftrightarrow ggg$)
in a unified manner. The back reaction channel is treated
fully consistently, exactly obeying  detailed balance.
The algorithm can handle, in principle,
any specified initial conditions
for the freed partons, the latter being on their kinetic mass shell.
First full simulations with minijet initial conditions 
demonstrate that the inclusion of the inelastic channels
leads to a very fast kinetic and chemical equilibration and also
to an early creation of pressure.
}
 
\begin{document}
 
\maketitle
\setcounter{page}{1}

\section{Introduction, Motivation and Summary}\label{intro}
The prime intention for present ultrarelativistic heavy ion collisions
at CERN and at Brookhaven lies in the possible experimental identification of
a new state of matter, 
the quark gluon plasma (QGP). Recent measurements \cite{Exp} at RHIC
of the elliptic
flow parameter $v_2$ for nearly central collisions suggest that - in comparison
to fits based on simple ideal hydrodynamical models - the
evolving system builds up a sufficiently  early
pressure and potentially also achieves (local) equilibrium.
On the other hand,   
the system in the reaction is at least initially
far from any (quasi-)equilibrium configuration.

To describe the dynamics of ultrarelativistic heavy ion collisions,
and to address the crucial question of thermalization and early
pressure buildup, we have developed a kinetic parton cascade algorithm
\cite{Xu04}
inspired by perturbative QCD including
for the first time inelastic (`Bremsstrahlung')
collisions $gg \leftrightarrow ggg $
besides the binary elastic
collisions. 
It is the aim to get a more detailed
and quantitative understanding of the early dynamical stages
of deconfined matter and to test various initial conditions
for the liberated gluons, than envoking ad hoc
phenomenological, hydrodynamical ansaetze.
It is well known, that a parton cascade analysis, incorporating only
elastic (and forward directed) $2 \leftrightarrow 2$
collisions described via one-gluon exchange,
shows that thermalization and early (quasi-)hydrodynamical behaviour
(for achieving sufficient elliptic flow)
can not be built up or maintained,
but only if a much higher cross section is being employed \cite{MG02}.
Hence, the collective flow phenomena observed at RHIC seem to indicate
that the early evolution of deconfined matter cannot be described
with standard pQCD inspired interactions, but can only be due
to much stronger and yet unknown interactions. This would suggest
that a QGP cannot be described via pQCD.
On the other hand, the possible importance of the inelastic reactions
on overall thermalization was
raised in the so called `bottom up thermalization' picture \cite{B01}.
It is intuitively clear that gluon multiplication should
not only lead to chemical equilibration, but
also should lead to a faster kinetic equilibration.
This represents one (but not all) important motivation
for developing a consistent algorithm to handle
also inelastic processes.

In the next section we briefly
detail on and show some numerical tests of this new scheme,
treating elastic and inelastic multiplication collisions
in a {\em unified} manner \cite{Xu04}.
Most importantly, the (multiparticle) back reation channel
($ggg\rightarrow gg$) is treated fully consistently
by respecting detailed balance within the {\em same} 
algorithm.
Incorporating this algorithm in a full
3+1 dimensional Monte Carlo cascade, we 
achieve a covariant parton cascade which can accurately handle
the immense elastic as well as inelastic scattering rates 
occuring inside the dense (gluonic) system.
Furthermore, we can then address
the question of the importantance of such (still) pQCD inspired reactions
on the thermalization and early
pressure build for heavy ion collisions at RHIC.
The algorithm can incorporate
any specified initial conditions
for the freed on-shell partons.
First results, taking as a conservative point of view 
minijet initial conditions, 
will be given in the last section:  
The exploratory study shows already 
that indeed the gluon multiplication
via Bremsstrahlung (and absorption) is of utmost importance,
where kinetic equilibration 
(for the present setting of initial conditions)
is achievd on a timescale of about 1 fm/c,
whereas the full chemical equilibration occurs on a somewhat
slower scale of about 2-3 fm/c.

\section{Treatment of elastic and inelastic multiparticle 
collisions in a unified scheme}\label{PCinel}  

\begin{figure}[ht]
\vspace*{0.1cm}
\includegraphics[width = 2.5in ]{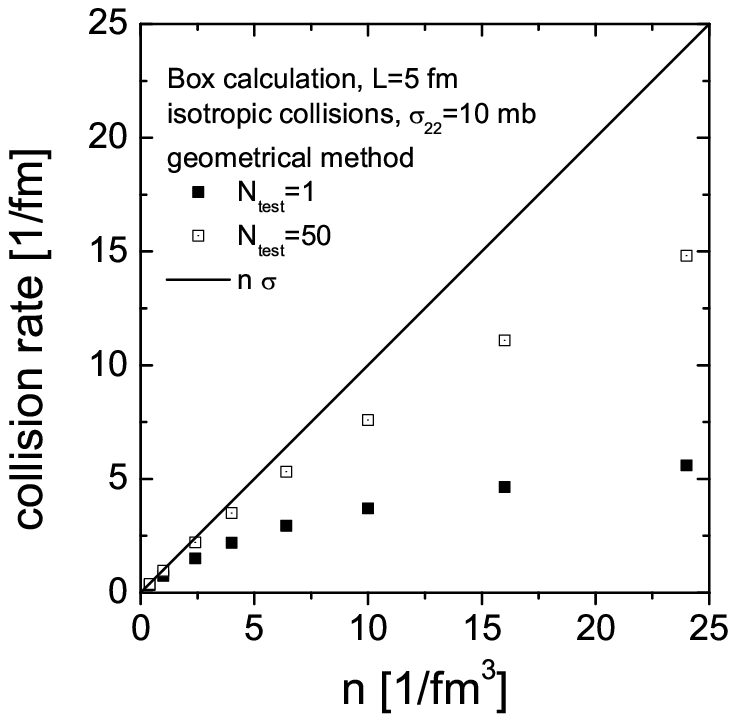}
\includegraphics[width = 2.5in ]{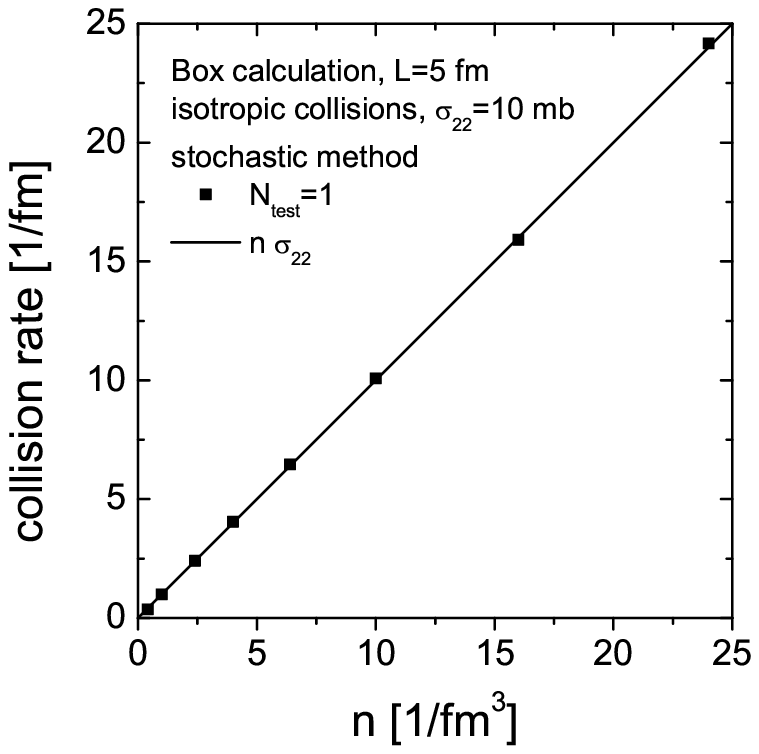}
\vspace*{-0.6cm}
\caption[]{
$2$ by $2$ elastic collision rate in a simulation of
propagating particles in a box versus the ideal rates
$1/(n\sigma ) $. The total cross section is fixed to 10 mb.
In the left (right) figure the geometrical (stochastic) method
is employed.}\label{Box1}
\end{figure}

In developing the algorithm special emphasis is put on
obeying the principle of detailed balance among the gain and loss
contributions.
The standard incorporation of $2 \leftrightarrow 2$
scattering processes in a transport
description is based on the geometric interpretation of the cross section
\cite{MG00}. For large particle densities,
however, such an implementation leads to
considerable problems to generate a causal collision sequence
among the various partons, resulting in a severe reduction of the collision
rate compared to the one dictated by the Boltzmann equation.
This numerical artefact is demonstrated in
left panel of 
fig.~\ref{Box1}.
In principle, this can be cured by the test particle method 
specifying each physical particle a specified number of
test particles (see left panel of  fig.~\ref{Box1} and also see \cite{MG00}).
However, for such a
situation of an ultra-dense plasma, where
the mean free path is in the order or comes below the
interaction length $\sqrt{\sigma / \pi }$, the stochastic method
(or `particle in cell' method, see eg \cite{La93}) is better suited to
solve the Boltzmann equation directly via transition rates in
sufficiently small spatial cells, see right panel of 
fig.~\ref{Box1}. The actual collision rate being realized in the
simulation
fully equals the ideal one of the Boltzmann equation which shall be
simulated. One can go to arbitrary high densities.
Moreover, this method can, in principle, be generalized to (any) multiparticle
scattering processes.

As outlined in the introduction, we want to incorporate
the direct Bremsstrahlungs process
$gg \leftrightarrow ggg$. Detailing on the the back reaction, we define
a transition probability in a time interval
with $0\leq P_{32} \ll 1$ for a given triplet of gluons
with specific momenta 
in a sufficiently small local cell:
\begin{equation}
\label{p32}
P_{32} = \frac{\Delta N_{coll}^{3\to 2}}{\Delta N_1 \Delta N_2 \Delta N_3}
= \frac{1}{8E_1 E_2 E_3} \frac{I_{32}}{N_{test}^2}
\frac{\Delta t}{(\Delta^3 x)^2} \, .
\end{equation}
Here $I_{32}$ is defined
as the to be calculated phase space integral 
$$
\frac{1}{2!} \int \frac{d^3 p^{'}_1}{(2\pi)^3 2E^{'}_1}
\frac{d^3 p^{'}_2}{(2\pi)^3 2E^{'}_2} | {\cal M}_{123\to 1'2'} |^2 (2\pi)^4
\delta^{(4)} (p_1+p_2+p_3-p^{'}_1-p^{'}_2) \, .
$$  
$\Delta^3 x $ denotes the volume of the specific cell and
$N_{test}$ represents the number of test-particles.
(The incorporation of test particles can also be applied, 
in addition, to achieve a better
statistics \cite{Xu04}.) The entering matrix element one obtains
from the $2\rightarrow 3$ element via a standard
prefactor governed by a detailed balance relation.
The latter is given by
\begin{equation}
\label{m23}
 | {\cal M}_{gg \to ggg} |^2 = ( \frac{9 g^4}{2} 
\frac{s^2}{({\bf q}_{\perp}^2+m_D^2)^2}  ) 
( \frac{12 g^2 {\bf q}_{\perp}^2}
{{\bf k}_{\perp}^2 [({\bf k}_{\perp}-{\bf q}_{\perp})^2+m_D^2]}  ) 
\theta (k_{\perp} \Lambda_g - \cosh y ) \,  ,
\end{equation}
where $g^2=4\pi\alpha_s$.
${\bf q}_{\perp}$ and ${\bf k}_{\perp}$ are the perpendicular component of
the momentum transfer and that of the momentum of the radiated gluon 
in the c.m.-frame of the collision, respectively.
$y$ denotes the rapidity of the
radiated gluon.
We thus consider
$gg \rightarrow ggg $
in leading-order of pQCD and
employ an effective Landau-Pomeranchuk-Migdal suppression
with $\Lambda_g$ denoting the gluon mean free
path, which is given by the inverse of the total gluon 
collision rate $\Lambda_g=1/R_g$,
and also employ a standard screening mass $m_D$
for the infrared sector of the
scattering amplitude.

\begin{figure}[ht]
\vspace*{0.1cm}
\includegraphics[width = 2.5in ]{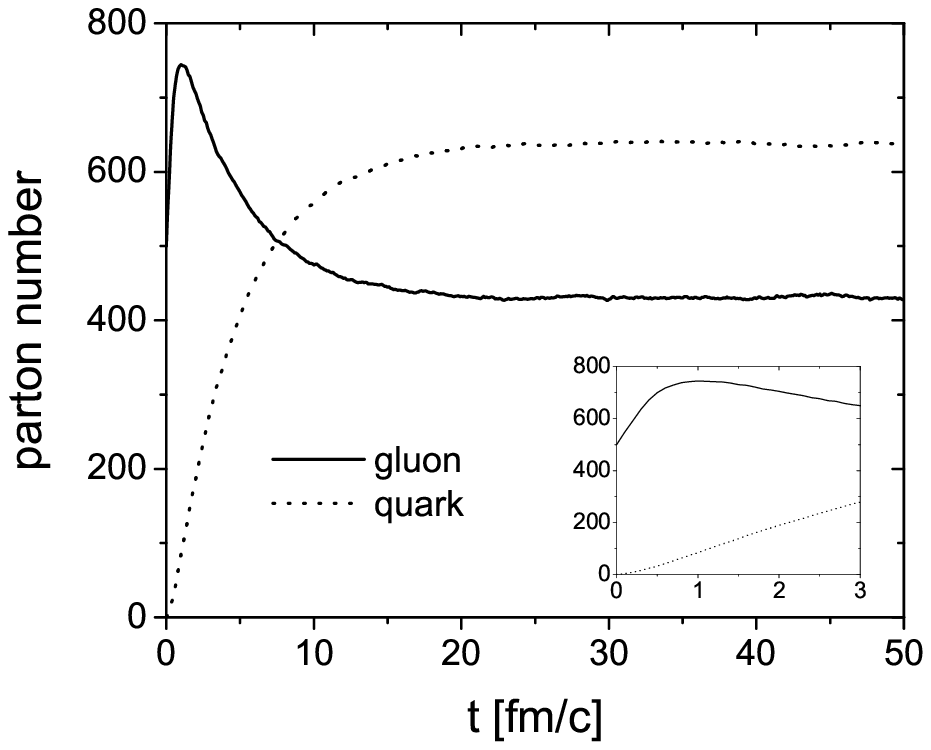}
\includegraphics[width = 2.7in ]{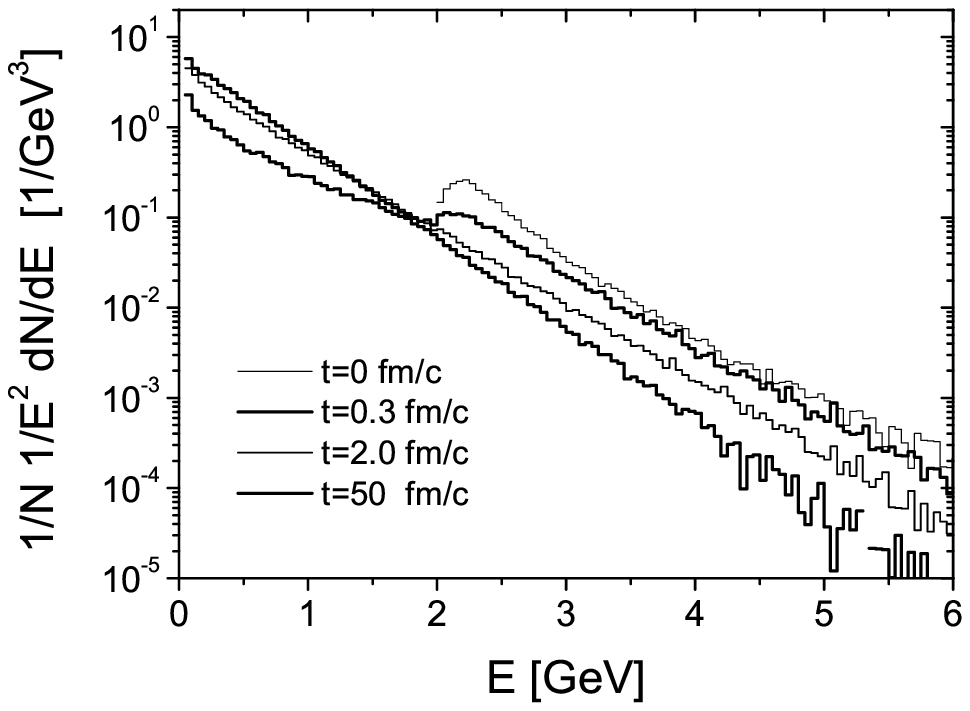}
\vspace*{-0.6cm}
\caption[]{Left panel: Time evolution of the gluon and 
quark numbers for a typical box calculation.
The gluon number starts at 500. Right panel:
Energy spectra at different times
within the box calculation.
At $t=0$ only minijets with $p_t>2$ GeV are populated.
Energy degradation to lower momenta proceeds rapidly by
gluon emission.
At $t\approx 0.5-1 $ fm/c full kinetic as well as chemical equilibrium
among the gluons is reached. The quark degrees of freedom do equilibrate
on a much slower time scale, roughly one order of magnitude larger.
}\label{Box2}
\end{figure}

To demonstrate the applicability of the method,
we show in the following the formation of a quark gluon plasma
within a fixed box and
study the way of kinetic and chemical equilibration for different parton
species. The initial conditions of the partons entering the cascade are
given by multiple minijet production from the binary 
nucleon-nucleon-scattering in a nucleus-nucleus-collision,
where we have chosen a conservatively large
transverse momentum cutoff of $p_t > p_{0} = 2$ GeV/c 
\cite{Es89},
according to the differential jet cross section:
\begin{equation}
\label{csjet}
\frac{d\sigma_{jet}}{dp_T^2dy_1dy_2} = K \sum_{a,b}
x_1f_a(x_1,p_T^2)x_2f_b(x_2,p_T^2) \frac{d\sigma_{ab}}{d\hat t} \, .
\end{equation} 
$p_T$ denotes the transverse momentum and $y_1$ and $y_2$ are the rapidities
of the produced partons.
The minijets are considered in the central rapidity interval $(-0.5:0.5)$
at RHIC energy of $\sqrt{s}=200$ GeV. The initial partons are distributed
uniformly in a cubic box of size 3 fm,
which approximately corresponds to the central region of heavy ion collions
at some early initial time.

Fig.~\ref{Box2} shows the time evolution of the gluon and quark number.
Gluon equilibration undergoes two stages: at first the gluon number
increases rapidly and then smoothly evolves to its final equilibrium value
together with the quark number at a much lower rate. 
The early stage of gluon production, on a timescale of
$0.5$ fm/c, also leads to an
immediate kinetic equilibration of the momentum distribution
(see also the right panel of fig.~\ref{Box2}),
as well as to a rather abrupt lowering of the
temperature by soft gluon emission until equal balance
among gain and loss contributions in the transitions is reached.
The later slower evolution is then governed by
chemical equilibration of the quark degrees of freedom.
The final temperature is identical
to the slope parameter
of the late energy spectra depicted in the right panel of fig.~\ref{Box2}.

\section{Thermalization in a real cascade operating at RHIC}
\label{PCtherm}

For implementing this scheme into a 3-D operating cascade,
the important task is to develop a dynamical mesh of
(expanding) cells in order to handle the extreme
initial situation \cite{Xu04}. For
the initial out-of-equilibrium conditions we take
minijets (\ref{csjet}) with a cutoff of 2 GeV and which
are distributed  in space-time via the corresponding overlap function.
The so produced number of initial gluons is about 900
with a $\frac{dN_g}{dy}\approx 200$ at midrapidity.
This is a rather low value, but we keep to this conservative 
estimate for our first exploratory study. 
In the following we also restrict ourselves
only to the dynamics of gluons. The coupling constant, the screening mass
as well as the LPM cutoff are calculated dynamically.

\begin{figure}[ht]
\vspace*{0.1cm}
\includegraphics[width = 2.6in ]{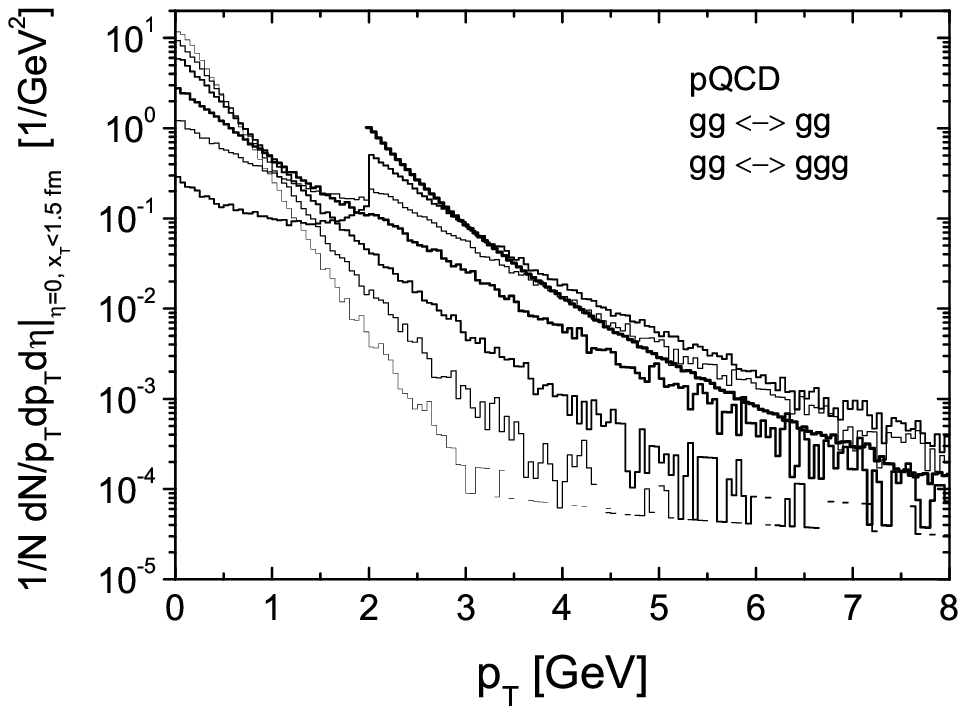}
\includegraphics[width = 2.6in ]{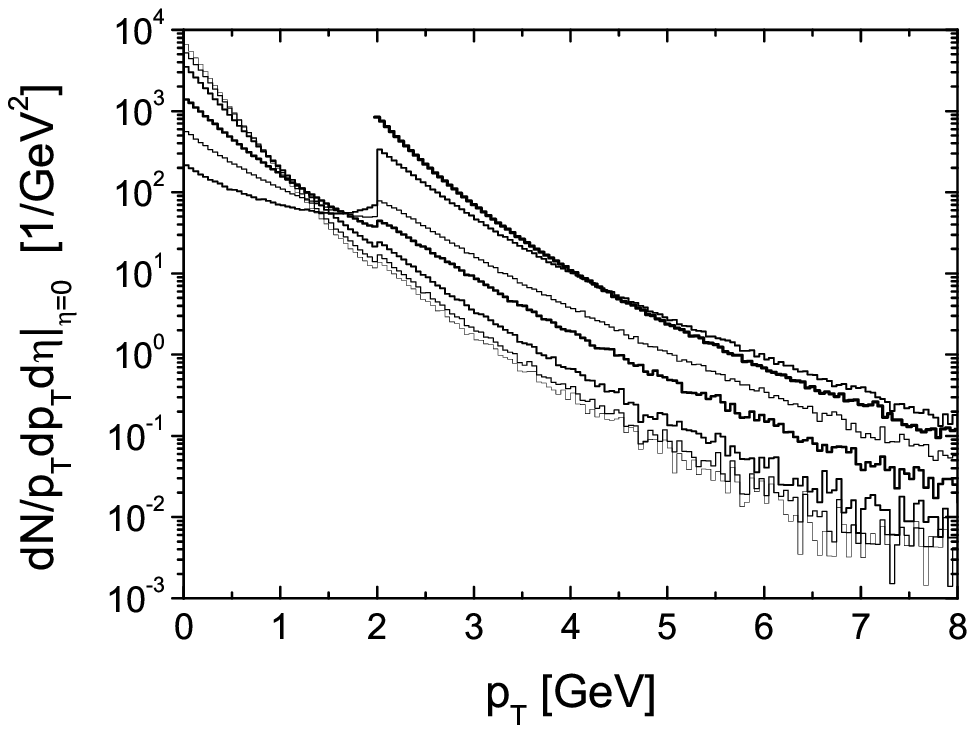}
\vspace*{-0.6cm}
\caption[]{Left panel:
Transverse momentum spectrum at
(spatial) midrapidity ($\Delta \eta =1$) at different times
(t=0.2, 0.5, 1, 2, 3 and 4 fm/c
from second upper to lowest line) 
for a real, central
fully 3-D 
ultrarelativistic heavy ion collision. Only the partons residing
in a central cylinder of radius $R\leq 1.5$ fm are depicted.
From $t=0$ on first only gluonic minijets with $p_t>2$ GeV are populated
(most-upper, boldfolded line).
Energy degradation to lower momenta proceeds rapidly by
gluon emission within the first fm/c. Maintenance of
(quasi-)kinetic and (later) chemical equilibrium is given up to 4 fm/c,
where longitudinal and transversal
(quasi-hydrodynamical) work is done resulting in a continous
lowering of the temperature.
 Right panel: Like the left panel, but now all partons in the transverse
direction are counted for. The evolution of the spectra 
at lower momenta is qualitatively similar to that of the central
cylinder. 
The initial non-equilibrium high momentum tail
following a standard power-law for mini-jet production is partly
surviving, stemming from the escaping mini-jets of the outer region.
}\label{dn}
\end{figure}
\begin{figure}[ht]
\vspace*{0.1cm}
\includegraphics[width = 2.6in ]{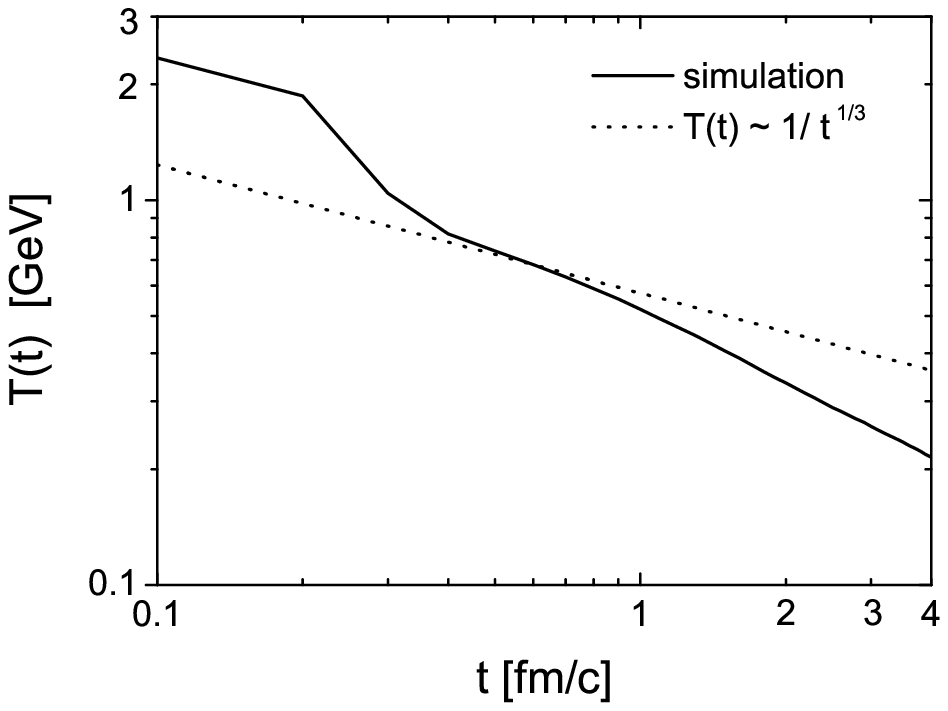}
\includegraphics[width = 2.6in ]{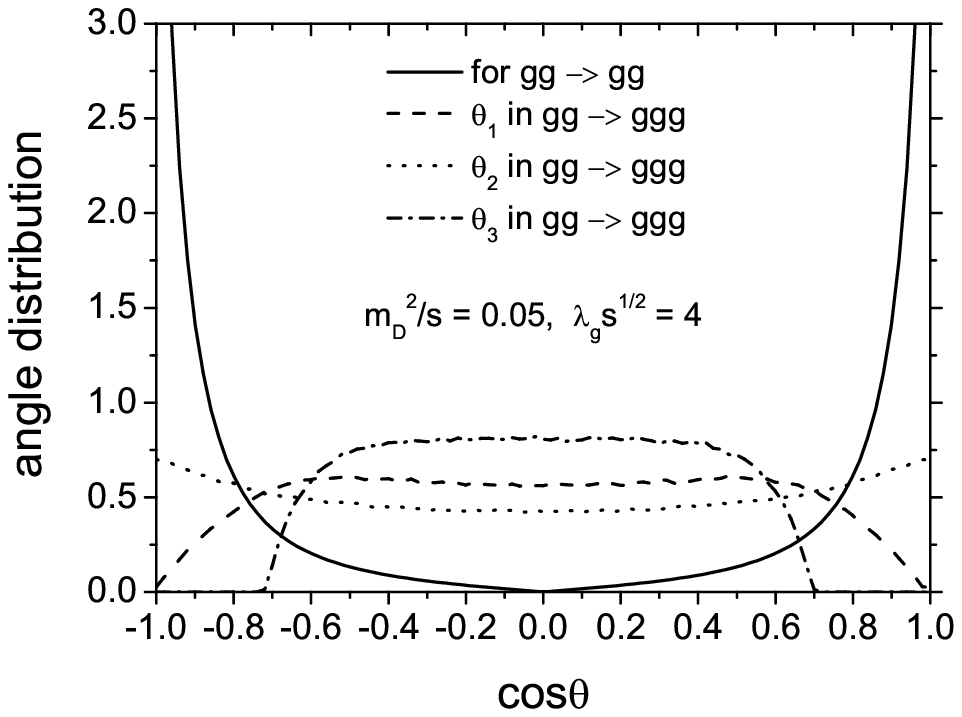}
\vspace*{-0.6cm}
\caption[]{Left panel:
The evolution of the effective temperature $T:=E/(3N)$ of the
expanding system in the central cylinder of the left panel of
Fig. \ref{dn} is shown.
After kinetic equilibration on a time-scale of 1 fm/c,
the further cooling proceeds on a noticeable
faster scale than suggested by one-dimensional ideal hydro-dynamical
expansion. This is mainly due to an ongoing production of
low momentum gluons, as the system is chemically saturated only
at times of about 3 fm/c.
Right panel: The angular distribution of the
scattering processes $gg\rightarrow gg$ and $gg \rightarrow ggg$
for a representative situation during the evolution.
Whereas the eleastic scattering is clearly forward peaked, this
is not the case for the inelastic reaction including a LPM cutoff.
Especially the emitted gluons show a flat and non-forward 
angle distribution.

}\label{fig4}
\end{figure}

In Figs. \ref{dn} we show the 
transverse momentum spectrum at
(spatial) midrapidity ($\Delta \eta =1$) at different times
for partons of a central cylinder of radius $R\leq 1.5$ 
and, respectively, of the total transversal region.
From $t=0$ on first only gluonic minijets with $p_t>2$ GeV are populated.
Energy degradation to lower momenta proceeds rapidly by
gluon emission within the first fm/c. Maintenance of
(quasi-)kinetic and (later) chemical equilibrium is given up to 4 fm/c,
where longitudinal and transversal
(quasi-hydrodynamical) work is done resulting in a continous
lowering of the temperature.
This can be seen by the continous steepening of the exponential
slopes of the spectrum with progressing times.
It turns out that kinetic momentum equilibration occurs at times
of about 1 fm/c, whereas full chemical equilibration occurs
on a smaller scale of about 3 fm/c. 
For the complete transversal region, of course,
a remedy of the initial non-equilibrium high momentum tail remains
stemming from the escaping minijets of the outer region.

In Fig. \ref{fig4}
we plot the evolution of the effective temperature $T:=E/(3N)$ of the
expanding system in the central cylinder. 
Indeed the further cooling after 1 fm/c proceeds on a noticeable
faster scale than suggested by one-dimensional Bjorken expansion.
This is mainly due to the ongoing production of
gluons. At the end of the evolution at t=4 fm/c about
$\frac{dN_g}{dy} \approx 400$ are at midrapidity, i.e. the gluon number has
doubled. If we compare the amount of transversal energy
at midrapidity with experimental factor, we are roughly below
by at least a factor of 2 to 3. This means, that in the initial conditions
too few gluons have been assumed. Indeed, there exist parametrizations
that the initial gluon number at midrapidity should be 
$\frac{dN_g}{dy} \approx 800$, i.e. a factor of 4 more than assumed here.
In turn, an incorporation of such a gluon number as initial conditions
would roughly shorten both the kinetic and chemical
equilibration by a factor of 4, so that a full thermalization should
be achieved on an accordingly smaller timescale! We leave this
for a future detailed investigation.
 
In the right panel of Fig.\ref{fig4}
the angular distribution of the
scattering processes $gg\rightarrow gg$ and $gg \rightarrow ggg$
for a representative situation during the evolution are shown.
Whereas the eleastic scattering is forward peaked, this
is not the case for the inelastic reaction including a LPM cutoff.
Especially the emitted gluons show a flat and non-forward 
angle distribution. This underlines why the inelastic processes
are so important not only for accounting for chemical equilibration,
but also for kinetic equilibration. It is these radiations that
actually brings about early thermalization to the QGP. 

In the future a lot of details have to be
explored: Thermalization (also of the
light and heavy quark degrees of freedom) has to be
investigated with
various initial conditions like minijets, 
with a detailed comparison to data,
or the color glass condensate,
serving as input for the so called 
bottom up picture \cite{B01}.
How likely is the latter picture for true coupling constants
and not parametrically small ones?
Furthermore we will study the impact parameter dependence
of the transverse energy in order to understand elliptic and transverse flow
at RHIC within this new scheme of a kinetic parton cascade including
pQCD inelastic interactions, again by comparing to experimental data.
Can the inelastic interactions generate the seen elliptic flow $v_2$? 
How good works (ideal) hydrodynamics?
Also the partonic jet-quenching picture can be analysed
in 3-D details.

\section*{Acknowledgement}
The work has been supported by BMBF, DFG and GSI Darmstadt.

\vfill\eject
\end{document}